\documentclass[%
 reprint,
superscriptaddress,
 amsmath,amssymb,
 aps,
]{revtex4-2}

\usepackage{graphicx}
\usepackage{dcolumn}
\usepackage{bm}
\usepackage{color}
\usepackage{multirow}
\usepackage{float}

\newcommand{\bks}[1]{\left( #1 \right)}
\newcommand{\squarebra}[1]{\left[ #1 \right]}
\newcommand{\curlybra}[1]{\left\{ #1 \right\}}
\newcommand{\rmcapsuper}[2]{{#1}^{\scriptscriptstyle \mathrm{#2}}}
\newcommand{\rmcapsub}[2]{{#1}_{\scriptscriptstyle \mathrm{#2}}}

\begin{document}

\preprint{APS/123-QED}

\title{Dielectric response and structural properties of finite-temperature electron liquids}

\author{Chengliang Lin}
 \affiliation{National Key Laboratory of Computational Physics, Institute of Applied Physics and Computational Mathematics, Beijing 100088, People’s Republic of China}
 
\author{Yong Hou}
 \email{yonghou@nudt.edu.cn}
\affiliation{College of Science, National University of Defense Technology, Changsha Hunan 410073, People’s Republic of China}

\author{Jianmin Yuan}
 \email{yuanjianmin@jlu.edu.cn}
 \affiliation{College of Science, National University of Defense Technology, Changsha Hunan 410073, People’s Republic of China}
\affiliation{Institute of Atomic and Molecular Physics, Jilin University, Changchun Jilin 130012, People’s Republic of China}

\author{Yong Wu}
 \email{wu\_yong@iapcm.ac.cn}
 \affiliation{National Key Laboratory of Computational Physics, Institute of Applied Physics and Computational Mathematics, Beijing 100088, People’s Republic of China}

\author{Jianguo Wang}
\affiliation{National Key Laboratory of Computational Physics, Institute of Applied Physics and Computational Mathematics, Beijing 100088, People’s Republic of China}


\begin{abstract}
The dielectric response and structural properties of finite-temperature electron liquids are central to accurately describing the physical behavior of electronic systems. This study presents a robust analytical model for the static structure factor of the uniform electron gas, combining physically motivated form for the static structure factor with constraints derived from high-accuracy path integral Monte Carlo simulations. The model accurately reproduces key features of the static structure factor across a broad range of temperatures and densities. Using this static structure factor, the density response function is directly evaluated, enabling a self-consistent definition of the static local field correction. As practical applications, the model is employed to investigate the low-velocity stopping power and the electron–ion friction coefficient. Results derived for the friction coefficient show good agreement with simulation data at moderate coupling and degeneracy. The proposed approach provides a computationally efficient and reliable method for characterizing the static response properties of correlated electron systems, facilitating improved simulations of energy deposition and ionic transport in warm dense matter and other strongly coupled quantum plasmas.
\end{abstract}

\maketitle


\section{\label{sec:intro}Introduction}
The uniform electron gas (UEG) serves as a fundamental model system for understanding correlated quantum many-body effects~\cite{Nozieres1999,Giuliani2008,Isihara1998,Zaghoo2019,Kim2022,Dornheim2025}. Understanding the behavior of UEG systems across varying densities and temperatures is crucial for accurately modeling warm dense matter~\cite{Glenzer2004,Brown2013,Dornheim2018,Kodanova2025}, inertial confinement fusion~\cite{He2018,Ding2018,Hayes2020,Hentschel2023,Hu2024}, and astrophysical environments~\cite{Chabrier1993,Vorberger2007,Kraus2017}. Despite its simplicity, the UEG exhibits rich physical behavior due to the interplay of Coulomb coupling, quantum degeneracy, and collective excitations, rendering traditional perturbative methods like the random phase approximation (RPA) inadequate~\cite{Ichimaru1987}. Thus, developing advanced analytical models, non-perturbative theoretical methods, and simulation approaches has become a major focus of current research.

The static structure factor (SSF) and the static density response function of UEG are central to characterizing the equilibrium properties of plasmas, such as the ionization potential depression~\cite{Lin2016,Lin2019,Zan2021} and the exchange-correlation functional~\cite{Utsumi1980,Karasiev2016,Groth2017,Hilleke2025}. It further provides key insights into electronic correlations, screening phenomena, and the onset of collective excitations. The SSF reveals how electrons are spatially correlated, offering a direct window into the organization of the quantum fluids, from ideal gases to strongly coupled liquids~\cite{Giuliani2008}. In the classic limit, the SSF is intimately connected to the static density response function according to the fluctuation-dissipation theorem~\cite{Dharma2000,Kremp2005}. In the strongly coupled and partially degenerate regimes where strong Coulomb interactions and quantum effects are equally important, accurately calculating these properties from theory becomes notoriously difficult. 

Unlike the classical one-component plasma (OCP), which can be described using static theories such as the Ornstein–Zernike equation~\cite{Ornstein1914,Baus1980,Hansen2006}, the SSF of UEG typically must be derived from its dynamical counterpart. This is primarily because the interplay of strong coupling and quantum effects as well as dynamic correlations complicates theoretical descriptions of the UEG. The simplest approximation for modeling the dynamical response within UEG is the RPA. To account for the influence of the strong Coulomb interactions and quantum effects, local field correction or collision frequency are introduced to improve the theoretical description based on the RPA~\cite{Hentschel2023,Fortmann2010,Dornheim2021,Kaehlert2024}. However, existing analytical models often do not satisfy the exact sum rules and asymptotic constraints~\cite{Atwal2002,Chuna2025}, which limit their predictive power and sometimes lead to unreasonable behavior of the correlation properties~\cite{Dornheim2020,Tanaka2016}. Recent advances in path-integral Monte Carlo (PIMC) simulations have enabled highly accurate calculations of properties of the UEG~\cite{Brown2013,Groth2019,Filinov2020,Larkin2021}, even in the strongly coupling regime, which can therefore serve as a testbed for theoretical methods. However, PIMC simulations are computationally prohibitive over wide parameter ranges. The purpose of this work is to present a robust analytical model for the SSF and the static density response function over a broad range of temperatures and densities.

In this work, we develop a comprehensive analytical framework for the SSF and static response function that is valid across a wide range of plasma conditions. Our strategy leverages highly accurate PIMC simulation data~\cite{Groth2017,Dornheim2020,Dornheim2017,Dornheim2020letter}, not as a direct fit, but as physical constraints to guide the development of an accurate model for the SSF. The resulting analytical expression successfully reproduces the key features of the SSF, including its characteristic peak and asymptotic limits, demonstrating consistent agreement with independent PIMC benchmarks. It also shows strong agreement with ab initio results for dielectric and static response properties in the plasma conditions under study. The availability of an accurate and computationally efficient model for the SSF and static response has immediate practical applications. A primary example is the calculation of the electron-ion friction coefficient~\cite{Zwicknagel2002,Montanari2017,Moldabekov2020}, which determines the low-velocity energy loss of a charged particle moving through an electron plasma. The accuracy of electron-ion friction coefficient critically affects the reliability of ion dynamics simulations, such as those based on the Langevin equation~\cite{Dai2010,Fu2018,Hou2021}. By providing a firm foundation in the well-established UEG model, our work facilitates more predictive and reliable simulations of energy deposition in complex, high-energy-density systems, and provides more reliable input parameters for future non-equilibrium dynamics simulations.

The present work is structured as follows: Sec.~\ref{sec:parameter} provides a brief introduction to key plasma parameters relevant for describing the UEG.
In Sec~\ref{sec:theory:ueg}, we outline the theoretical framework employed in our calculations and discuss the relationships among the SSF, the static response function, and the static local field correction. The numerical strategy used to calculate the SSF of the UEG is detailed in Sec.~\ref{sec:theory:ssf}. Sec.~\ref{sec:theory:friction} presents the theory underlying the low-velocity stopping power and its connection to the electron-ion friction coefficient. Our results for the SSF, the static response function, and the friction coefficients are presented and discussed in Sec.~\ref{sec:results}. Finally, conclusions are summarized in Sec.~\ref{sec:conclusions}.

\section{\label{sec:theory}Theoretical method}
\subsection{\label{sec:parameter}Parameters for Coulomb coupling and electron degeneracy}
The UEG is a quantum many-body system consisting of free electrons embedded in a uniform neutralizing background of positive charges. It can be viewed as the quantum-mechanical analogue of the classical OCP system. In this work, we consider a homogeneous paramagnetic electron liquid with electron number density $n_\mathrm{e}$ and thermal temperature $T$. While the classical OCP is fully characterized by a single coupling parameter $\Gamma = e^2 / (4 \pi \epsilon_0 a_\mathrm{ee} \rmcapsub{k}{B} T)$~\cite{Hansen1973}, where $a_\mathrm{ee} = (4 \pi  n_\mathrm{e} / 3)^{-1/3}$ denotes the average electron-electron separation, the UEG requires two independent dimensionless parameters for its complete description. Although different choices of these two parameters exist in the literature~\cite{Dornheim2018,Ichimaru1987,Brown2013,Tanaka2017,Dandrea1986}, we adopt the following pair of parameters to describe the UEG: the electron degeneracy parameter $\theta$ and the Brueckner parameter $\rmcapsub{r}{S}$
\begin{equation}
    \theta = \frac{\rmcapsub{k}{B} T}{\rmcapsub{E}{F}}, \quad \rmcapsub{r}{S} = \frac{a_\mathrm{ee}}{\rmcapsub{a}{B}},
\end{equation}
where $\rmcapsub{a}{B}$ is the Bohr radius and $\rmcapsub{E}{F} = \hbar^2 \rmcapsub{k}{F}^2 / (2 m_\mathrm{e})$ denotes the Fermi energy with $\rmcapsub{k}{F} = \bks{3 \pi^2 n_\mathrm{e}}^{1/3}$ being the Fermi wavevector. The Brueckner parameter $\rmcapsub{r}{S}$ quantifies the Coulomb coupling strength in degenerate electron systems. Other dimensionless parameters can be expressed in terms of these two parameters. For example, the classical coupling parameter is given by $\Gamma = 2 \gamma_0^2 \, \rmcapsub{r}{S} / \theta$ with $\gamma_0 = (9 \pi / 4)^{-1/3}$. 

The quantum coupling strength $\rmcapsub{r}{S}$ and the classical coupling parameter $\Gamma$ can be unified within a single expression~\cite{Dandrea1986}. By definition, the Coulomb coupling strength is determined by the ratio of the average potential energy $E_\mathrm{pot} = \frac{e^2}{4 \pi \epsilon_0 a_\mathrm{ee}}$ and the mean thermodynamic kinetic energy $E_\mathrm{kin} = \langle \frac{m_\mathrm{e} v^2 }{2}\rangle$
\begin{equation}
    \Gamma_\mathrm{eff} = \frac{3}{2}\frac{E_\mathrm{pot}}{E_\mathrm{kin}} = \frac{2 \gamma_0^2 \, \rmcapsub{r}{S}}{\theta^{5/2} \, F_{3/2}(\eta)} =
    \begin{cases}
        5 \gamma_0^2 \rmcapsub{r}{S}, & \theta \ll 1 \\
        \Gamma, & \theta \gg 1
    \end{cases}
\end{equation}
where the factor $3/2$ is introduced to recover the classic coupling parameter in the high-temperature limit. The mean thermodynamic kinetic energy is given by $E_\mathrm{kin} = \frac{3}{2} \rmcapsub{E}{F} \theta^{5/2} \, F_{3/2}(\eta)$ with the Fermi–Dirac integral of order $\nu$~\cite{Arista1981,Arista1984}
\begin{equation}
    F_\nu(\eta) = \int_0^\infty \frac{x^\nu}{e^{x-\eta} + 1} dx .
\end{equation}
Here $\eta$ represents the reduced chemical potential of the electrons, determined implicitly by the condition $F_{1/2}(\eta) = 2 / (3 \theta^{3/2})$. Throughout this work, we examine the SSF and the static density response function within the parameter range  $1 \leq \rmcapsub{r}{S} \leq 100$ and $\theta \geq 0.5$.

\subsection{\label{sec:theory:ueg}Dielectric formulation and density response}
To investigate interparticle correlations in the UEG, we adopt the dielectric formulation in this work. Within this framework, the dynamic structure factor of free electrons $\mathbb{S}(\mathbf{k}, \mathrm{w})$ can be expressed in terms of the inverse dielectric function $\mathrm{Im} \varepsilon^{-1}(\mathbf{k}, \mathrm{w})$ dependent on the wave vector $\mathbf{k}$ and frequency $\mathrm{w}$ using the fluctuation-dissipation theorem~\cite{Kremp2005}
\begin{equation}
    \mathbb{S}(\mathbf{k}, \mathrm{w}) = -\frac{\hbar}{\pi n_\mathrm{e} \rmcapsub{v}{C}(k)} \mathrm{Im} \squarebra{\frac{1 + \rmcapsub{n}{B}(\mathrm{w})}{\varepsilon(k, \mathrm{w})}} ,
\end{equation}
where $\rmcapsub{v}{C}(k) =e^2 / (\epsilon_0 k^2)$ is the Coulomb potential in momentum space, and $\rmcapsub{n}{B}(\omega) = [e^{\beta \hbar \omega} - 1]^{-1}$ is the Bose distribution function with $\beta = (\rmcapsub{k}{B} T)^{-1}$. It is convenient to introduce the dimensionless wavenumber $q = k \, a_\mathrm{ee}$ and normalized frequency $\omega  = \mathrm{w} / \omega_\mathrm{pl}$, where $\omega_\mathrm{pl} = \sqrt{e^2 n_\mathrm{e} / (\epsilon_0 m_\mathrm{e})}$ is the plasma frequency of electrons. The dynamic structure factor can then be written in normalized form as
\begin{equation}
    S(q, \omega) = \omega_\mathrm{pl} \mathbb{S}(k, \mathrm{w})  = \frac{q^2}{3 \pi \Gamma} B\bks{d_\mathrm{e} \omega} \mathcal{L}(q, \omega)
\end{equation}
with $d_\mathrm{e} = \beta \hbar \omega_\mathrm{pl} = \gamma_0^2 \sqrt{12 \rmcapsub{r}{S}} / \theta$ and $B(x) = x [1-e^{-x}]^{-1}$. Here, $\mathcal{L}(q, \omega) = - \mathrm{Im} \varepsilon^{-1}(q, \omega) / \omega$ is the loss function, which is a positively defined quantity that characterizes the excitation spectrum and the stopping power in plasmas~\cite{Arista1984}. The SSF $S(q)$ is obtained by integrating $S(q, w)$ with respect to the frequency $\omega$
\begin{equation}\label{dsf2ssf}
    S(q) = \int_{-\infty}^\infty  S(q, \omega) \, d \omega .
\end{equation}

The central quantity characterizing the dynamical structure and response properties is the (inverse) dielectric function, which is highly non-trivial to be determined accurately. A well-known approximation is the RPA dielectric function~\cite{Arista1984}, which describes the effective response in a non-interacting electron gas. Various extensions beyond RPA have been developed, such as those incorporating a local field correction and the Born–Mermin ansatz~\cite{Fortmann2010}. In the present work, the inverse dielectric function is constructed within the framework of the canonical solution to the Hamburger truncated moment problem~\cite{Ara2021}, modeled as
\begin{equation}\label{lossfunc}
    \mathcal{L}(q, \omega) \! = \! \pi \squarebra{\bks{\frac{1}{\omega_1^2} \! - \! \frac{1}{\omega_2^2}} \delta(\omega) \! + \! \frac{1}{\omega_2^2} \delta\bks{\omega^2 \! - \! \omega_2^2}},
\end{equation}
where the characteristic frequencies $w_1$ and $w_2$ are defined by 
\begin{equation}
    \omega_1 \! = \! \omega_1(q) \! = \! \sqrt{\frac{C_2}{\omega_\mathrm{pl}^2 \, C_0(q)}}, \quad
    \omega_2 \! = \! \omega_2(q) \! = \! \sqrt{\frac{C_4(q)}{\omega_\mathrm{pl}^2 \,  C_2}}.
\end{equation}
Here, $C_\nu$ denotes the frequency moment of the loss function
\begin{equation}
    C_\nu \! = \! \frac{\omega_\mathrm{pl}^\nu}{\pi} \int_{-\infty}^\infty \omega^\nu \mathcal{L}(q, \omega) d\omega ,
\end{equation}
which is related to the sum rules (conservation laws) of the UEG. For instance, $C_2 = \omega_\mathrm{pl}^2$ corresponds to the $f$-sum rule reflecting the conservation of the particle number. The quantity $C_0(q)$ is related to the static inverse dielectric function $\varepsilon^{-1}(q) = \varepsilon^{-1}(q, 0)$ through the Kramers–Kronig relation~\cite{Ara2021}
\begin{equation}
    C_0(q) = 1 - \varepsilon^{-1}(q).
\end{equation}
For the OCP system, the function $C_4(q)$ is determined by the following expression
\begin{equation}
    C_4(q) = \omega_\mathrm{pl}^4 \squarebra{1 + K(q) + U(q)}
\end{equation}
with the function $U(q)$ defined as
\begin{equation}
    U(q) = \frac{1}{3 \pi} \int_0^\infty dp\, p^2 f(p,q) \squarebra{S(p) - 1},
\end{equation}
in which $f(p,q)$ is a monotonically decreasing function that depends only on the ratio $p/q$
\begin{eqnarray}
    f(p,q) = \frac{5}{6} - \frac{p^2}{2 q^2} + \frac{q}{4 p} \bks{\frac{p^2}{q^2} - 1}^2 \ln \left\vert \frac{q + p}{q-p} \right\vert .
\end{eqnarray}
The function $K(q)$ is expressed as
\begin{equation}\label{kqfunc}
    K(q) = A(q) \, \frac{q^2}{\Gamma} \theta^{3/2} F_{3/2}(\eta) + \frac{q^4}{12 \, \rmcapsub{r}{S}},
\end{equation}
Although $A(q) = 1$ was used in Ref.~\cite{Ara2021}, it is known that the $q^2$-term represents the exact kinetic energy per electron in the interacting UEG~\cite{Dornheim2018, Kugler1970}. Therefore, $A(q)$ should correspond to the ratio of the kinetic energy per electron in the interacting UEG to that in the non-interacting system, and generally differs from unity. In the current work, $A(q)$ is treated as a parameterization function chosen to ensure the correct asymptotic behavior of the static density response function in both the short- and long-wavelength limits.

Within the framework of the canonical solution to the Hamburger truncated moment problem (Eq. \eqref{lossfunc}), the static inverse dielectric function $\varepsilon^{-1}(q))$ can be directly obtained using the SSF as follows~\cite{Ara2021}
\begin{equation}
    \varepsilon^{-1}(q) = 1 - \frac{3\Gamma}{q^2} S(q) - \frac{1}{w_2^2} + \frac{d_\mathrm{e}}{2 w_2} \mathrm{coth}\bks{\frac{d_\mathrm{e} w_2}{2}} .
\end{equation}
The static density response function $\chi(q) $ is connected to the static inverse DF $\varepsilon^{-1}(q)$ through the relation 
\begin{align}
    \chi(q) = \frac{1}{\rmcapsub{v}{C}(q)} \squarebra{\varepsilon^{-1}(q) - 1} 
\end{align}
The availability of $\chi(q)$ further enables the determination of the static local field correction
\begin{equation}\label{gqslfc}
    G(q) = 1 - \frac{1}{\rmcapsub{v}{C}(q)} \squarebra{\frac{1}{\chi_0(q)} - \frac{1}{\chi(q)}}
\end{equation}
in which $\chi_0(q)$ denotes the static density response function of the ideal UEG and is given by~\cite{Arista1984}
\begin{equation}
    \rmcapsub{v}{C}(q) \chi_0(q) = - \frac{\gamma_0 \rmcapsub{r}{S}}{2 \pi z^3} g(z)
\end{equation}
with $z = k / (2 \rmcapsub{k}{F}) = \gamma_0 q/2$, and $g(z)$ given by
\begin{equation}\label{gzfunc}
    g(z) = \int_0^\infty \frac{y}{e^{y^2/\theta - \eta} + 1} \ln \left\vert \frac{z+y}{z-y} \right\vert \, dy .
\end{equation}

The electronic cusp condition~\cite{Kimball1973} predicts that the static local field correction approaches a constant in the large $q$ limit as $G(q \rightarrow \infty) = 1 - g(0)$, where $g(0)$ is the on-top value of the radial distribution function. Using the short-wavelength asymptotic form of $\chi_0(q)$ from Eq.~\eqref{gzfunc}, namely $g(z \gg 1) \approx 2/(3x) + \theta^{5/2} F_{3/2}(\eta) / (3 x^3)$, the asymptotic behavior of both $\chi(q)$ and $\omega_2(q)$ in the large-$q$ regime can be derived. A detailed analysis indicates that the function $A(q)$ in Eq.~\eqref{kqfunc}, which contributes to $K(q)$, must vanish in this limit. Accordingly, we propose the following form for $A(q)$
\begin{equation}\label{aqfunc}
    A(q) = \frac{1}{2} - \frac{1}{2} \tanh[0.1325 \, (q - 10.3725)] .
\end{equation}

\subsection{\label{sec:theory:ssf}Static structure factor}
In principle, the SSF of UEG $S(q)$ must be derived from its dynamic counterpart $S(q, w)$, since the quantum nature and the degeneracy effects of electrons complicate the development of statistical theories analogous to the classical Ornstein–Zernike equation. For classical OCP systems, Bretonnet and Derouiche proposed a simple analytic representation for the SSF~\cite{Bretonnet1988}. Inspired by its success and simplicity~\cite{Bretonnet1988,Veysman2016}, we adopt a similar analytical form for the SSF of UEG
\begin{equation}
    \frac{1}{S(q)} \! = \! 1 \! - \! \frac{3 \alpha_0}{q^4 \alpha_2^2} \! \! \squarebra{\cos\bks{q \alpha_1} \! + \! 2 \cos\bks{q \alpha_2} \! - \! 3 \frac{\sin\bks{q \alpha_1}}{q \alpha_1} \! }
\end{equation}
In the small-wavenumber limit ($q \rightarrow 0$), this form of SSF satisfies the exact relation $S(q \rightarrow 0) = q^2 / \alpha_0$. In classical OCP systems, this corresponds to the perfect screening condition, implying $\alpha_0 = \Gamma$. For the UEG, however, the exact long-wavelength limit of the SSF that is valid at any temperature and degeneracy is given by the following exact relation~\cite{Kugler1970}
\begin{equation}
    S(q \rightarrow 0) = \frac{q^2 }{\sqrt{12 \, \rmcapsub{r}{S}}} \, \mathrm{coth}\bks{ \frac{\sqrt{3 \rmcapsub{r}{S}} \, \gamma_0^2}{\theta}}.
\end{equation}
This relation leads to $\alpha_0 = \sqrt{4 \, \rmcapsub{r}{S}/3} \, \mathrm{tanh}(\sqrt{3 \rmcapsub{r}{S}} \, \gamma_0^2 / \theta)$. To fully determine the SSF, the parameters $\alpha_1$ and $\alpha_2$ must be specified for given temperature and density conditions. These can be obtained, for instance, by fitting to structure factors computed from alternative methods such as PIMC simulations. In the current work, the parameters $\alpha_1$ and $\alpha_2$ are determined by imposing physical constraints based on the excess internal energy and the on-top value of the radial distribution function.

The on-top value of the radial distribution function is given by
\begin{equation}\label{g0func}
    g(0) = 1 + \frac{2}{3 \pi} \int_0^\infty dq \, q^2 \squarebra{S(q) - 1}
\end{equation}
We require that $g(0) \equiv \rmcapsub{g}{PIMC}(0)$, where $\rmcapsub{g}{PIMC}(0)$ represents the value obtained from the PIMC-based fitting model~\cite{Dornheim2020letter}. The excess internal energy per electron is evaluated by
\begin{equation}\label{uexfunc}
    u_\mathrm{ex} = \frac{\rmcapsub{E}{H}}{\pi \, \rmcapsub{r}{S}} \int_0^\infty dq \, \squarebra{S(q) - 1},
\end{equation}
where $\rmcapsub{E}{H}$ is the Hartree energy. The second constraint is to minimize the absolute relative difference between the calculated $u_\mathrm{ex}$ and the PIMC result~\cite{Groth2017,Dornheim2018}, i.e., $\mathrm{min} [| u_\mathrm{ex} / \rmcapsuper{u}{PIMC}_\mathrm{ex} - 1 |]$. In the present approach, the parameters $\alpha_1$ and $\alpha_2$, and hence the resulting SSF, depend critically on the precision and validity of $g(0)$ and $u_\mathrm{ex}$ obtained from PIMC simulations. Under certain parameter conditions, exact agreement between the values computed from Eqs.~\eqref{g0func} and~\eqref{uexfunc} and the PIMC benchmarks cannot be achieved. Therefore, the values of $\alpha_1$ and $\alpha_2$ are determined self-consistently through a numerical minimization procedure.

\subsection{\label{sec:theory:friction}Low-velocity stopping power and electron-ion friction coefficient}

In the study of charged particle interactions with the UEG and plasmas, the stopping power $S(v) = - dE/dx$ plays a undamental role in characterizing energy deposition processes. Within the framework of linear response theory, the stopping power of a test ion with charge $Ze$ and velocity $v$ in the UEG system is given by~\cite{Arista1981,Kremp2005}
\begin{eqnarray}\label{stopping}
    S(v) = \frac{2 Z^2 e^2}{4 \pi^2 \epsilon_0 v^2} \int_0^\infty \frac{dk}{k} \int_0^{kv} d \mathrm{w} \, \mathrm{w} \, \mathrm{Im} \squarebra{\frac{-1}{\varepsilon(k, \mathrm{w})}},
\end{eqnarray}
where the recoil of the heavy ion due to collisions with electrons is neglected, corresponding to the assumption of an infinite test ion mass. In the low-velocity limit $v \ll v_\mathrm{th}$, where $v_\mathrm{th} = \sqrt{\rmcapsub{k}{B} T/m_\mathrm{e}}$ is the electron thermal velocity, the stopping power exhibits a linear dependence on the projectile velocity~\cite{Arista1981,Kremp2005,Tanaka1985}. This behavior motivates the introduction of the electron-ion friction coefficient $\xi = \frac{S(v)}{v}\vert_{v=0}$~\cite{Kremp2005,Zwicknagel2002}.

\begin{figure*}[t]
\includegraphics[width=0.9 \textwidth]{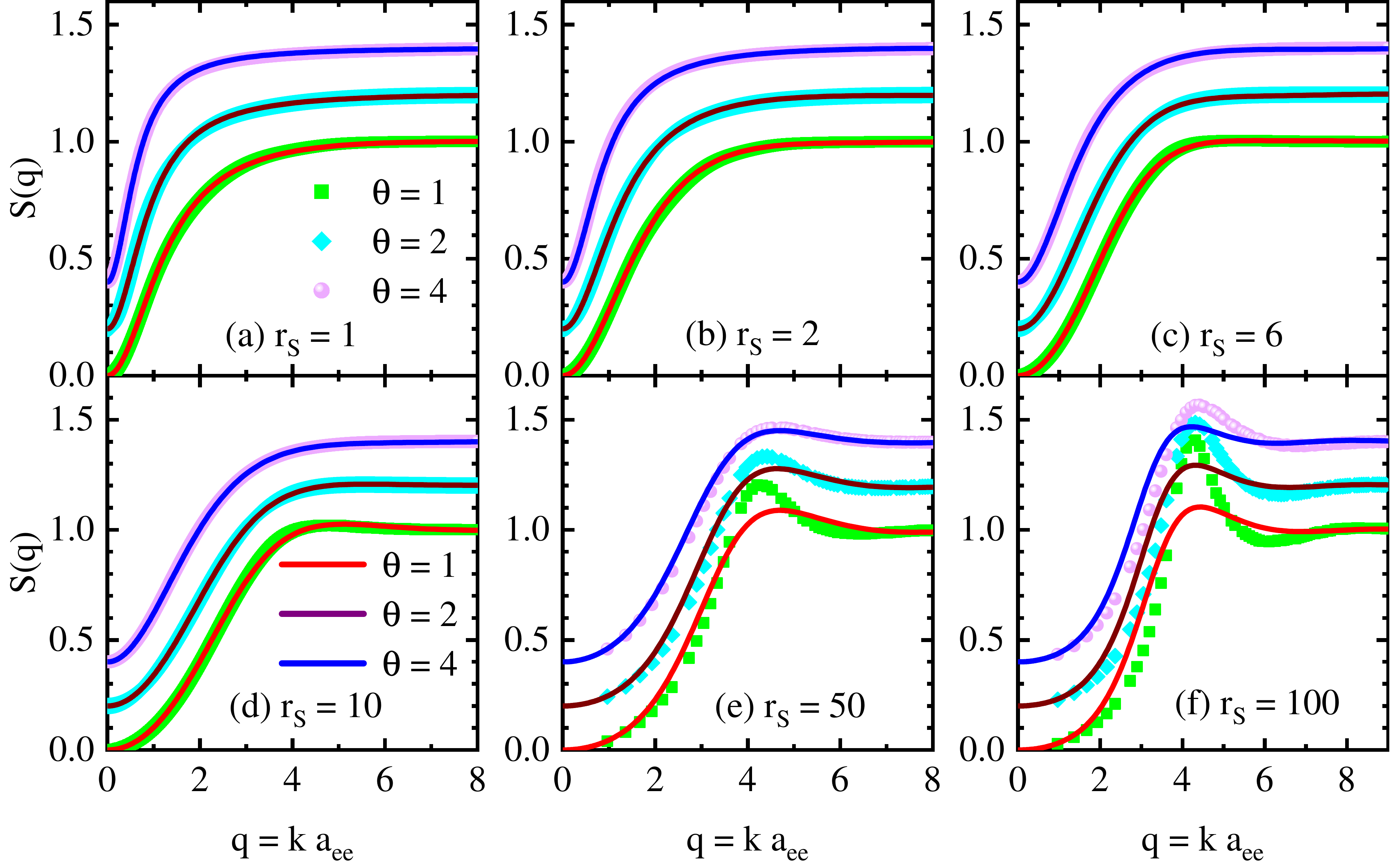}
\caption{\label{fig:sfg1g2}Comparison of the SSF for UEG under different plasma conditions. The square, diamond, and circular symbols represent the PIMC results~\cite{Dornheim2017,Dornheim2020} for three degeneracy parameters $\theta=1, \theta = 2, \textrm{and } \theta = 4$ at different densities, respectively. The corresponding predictions calculated from the model proposed in this work is depicted by the red $(\theta=1)$, brown $(\theta=2)$ and blue $(\theta=4)$ lines.}
\end{figure*}

The accurate determination of the electron–ion friction coefficient $\xi$ depends critically on a precise description of the dynamic density response, particularly the behavior of the local field correction in the static and long-wavelength limits. In principle, the dielectric (loss) function~\eqref{lossfunc} could be used to calculate the low-velocity stopping power. It is believed that the dominant contribution to the stopping power arise from single-particle excitations~\cite{Bringa1996,Archubi2022}. However, the loss function defined in Eq.~\eqref{lossfunc} fails to capture this essential physics and yields unphysical results. To overcome this limitation, non-canonical solutions to the Hamburger moment problem may be applied~\cite{Arkhipov2019}. In this work, we instead calculate the inverse dielectric function using the static local field correction provided in Eq.~\eqref{gqslfc}
\begin{equation}
    \frac{1}{\varepsilon(k, \mathrm{w})} = 1 + \frac{\rmcapsub{v}{C}(k) \chi_0(k, \mathrm{w})}{1 - \rmcapsub{v}{C}(k) [1 - G(k)] \chi_0(k, \mathrm{w})}.
\end{equation}
Substituting this expression into the stopping power formula~\eqref{stopping} and taking the low-velocity limit yields the following expression for the friction coefficient $\xi$
\begin{equation}\label{frictionlfc}
    \xi = \xi_0 \frac{4\gamma_0 Z^2 \rmcapsub{r}{S}^2}{3 \pi} \sqrt{\frac{2}{\theta}} \int_0^\infty dz \frac{z^3 \, F(z)}{(1 + e^{z^2/\theta-\eta})},
\end{equation}
with $\xi_0 = m_\mathrm{e} v_\mathrm{th} / a_\mathrm{ee}$ and the function $F(z)$ defined as
\begin{eqnarray}\label{frictionlfcFfunc}
    F(z) = \curlybra{z^2 + \frac{\gamma_0 \rmcapsub{r}{S}}{2 \pi z } g(z) [1 - G(z)] }^{-2}.
\end{eqnarray}
Setting $G(z) = 0$ recovers the well-known RPA result for the low-velocity stopping power and the corresponding friction coefficient. The friction coefficient $\xi$ is directly related to the quantity $\gamma = \xi/m_\mathrm{i}$ that appears in the Langevin equation of motion~\cite{Dai2010,Fu2018,Hou2021}, governing the dynamics of ion with mass $m_\mathrm{i}$ in an electron fluid.

\section{\label{sec:results}Results and discussion}

\subsection{\label{sec:results:structure}Parametrization of the static structure factor}

\begin{figure*}[t]
\includegraphics[width=0.9 \textwidth]{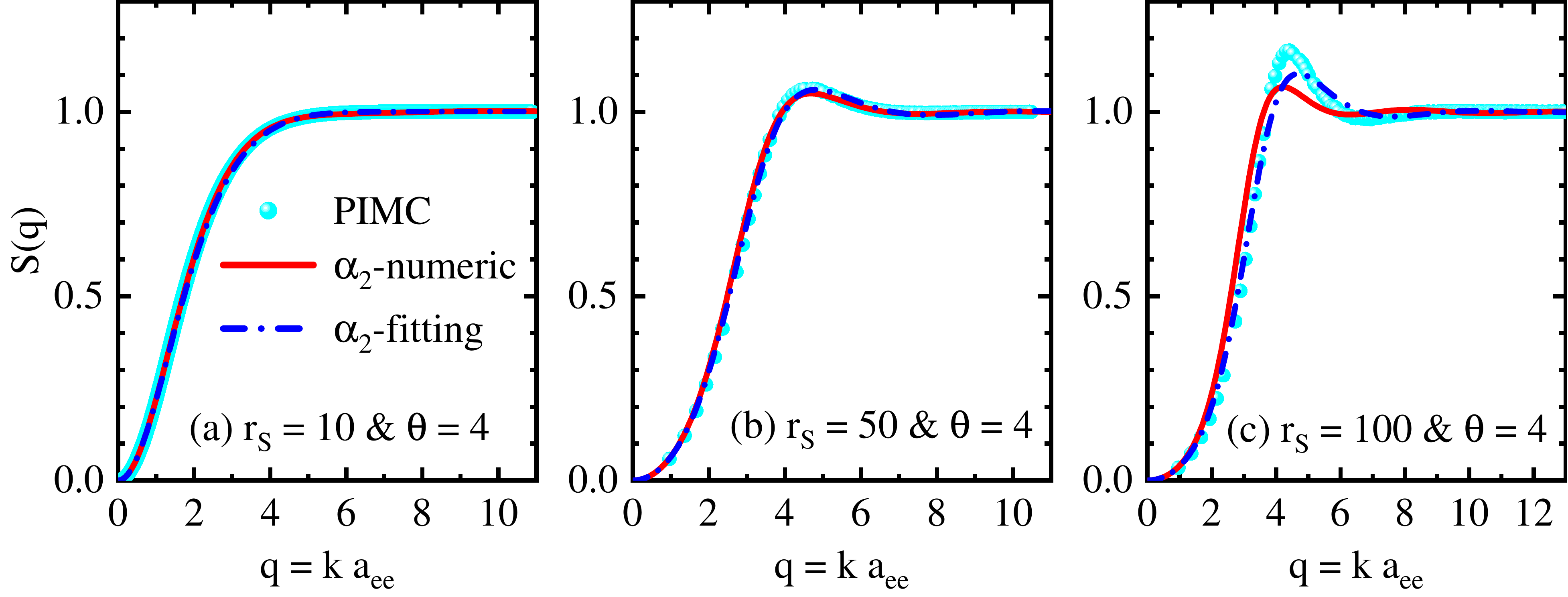}
\caption{\label{fig:sfg2fix}Comparison of the SSF for UEG under different plasma conditions. The square symbols represent the PIMC results~\cite{Dornheim2017,Dornheim2020} for three densities $\rmcapsub{r}{S} = 10$ (panel (a)),  $\rmcapsub{r}{S} = 50$ (panel (b)), and $\rmcapsub{r}{S} = 100$ (panel (c)) at the fixed degeneracy $\theta = 4$. The red lines mark the prediction obtained with the numeric strategy described in Sec.~\ref{sec:theory:ssf}. The blue curves represent the results using the fitting formula~\eqref{diameter}.}
\end{figure*}

In this section, we compare the SSF of UEG under various plasma conditions with the corresponding PIMC results~\cite{Dornheim2017,Dornheim2020}. The comparisons are summarized in Fig.~\ref{fig:sfg1g2}. As shown in the figure, the SSF model developed in this work performs exceptionally well in the parameter region of $1 \le \rmcapsub{r}{S} \le 10$ and $\theta \ge 1$, where it reproduces the PIMC data with high precision. At $\rmcapsub{r}{S} = 10$ and $\theta = 1$ ($\Gamma = 5.43$), the PIMC results exhibit a pronounced correlation peak, signaling the emergence of strong interparticle correlations. At an even higher coupling strength at $\rmcapsub{r}{S} = 50$ and $\theta = 4$ ($\Gamma = 6.79$), the electron correlations are further enhanced, resulting in a more distinct correlation peak. Remarkably, under these conditions, our model still shows overall agreement with the PIMC benchmarks, particularly in matching the position and height of the correlation peak. As $\theta$ decreases at a certain fixed density (such as at $\rmcapsub{r}{S}=50$), the coupling strength increases significantly, and noticeable discrepancies emerge between the SSF predicted by our model and the PIMC reference data.

Moreover, under low-temperature and low-density conditions (panel (f) in Fig.~\ref{fig:sfg1g2}), the proposed SSF model yields a single-peak structure, in contrast to the oscillatory behavior observed in PIMC simulations. It should be noted that in classical OCP systems, the model introduced by Bretonnet and Derouiche~\cite{Bretonnet1988} successfully reproduces oscillatory features consistent with those observed in metallic systems. These observations emphasize the need for further investigation into the combined effects of Coulomb coupling and electron degeneracy to achieve a more accurate description of the UEG SSF. Furthermore, there is currently no consensus on the SSF of electron fluids in the low-temperature and low-density region, where the results of different theoretical models exhibit substantial discrepancies~\cite{Tolias2023}. These limitations highlight the demand for more advanced theoretical frameworks capable of simultaneously capturing strong correlations and quantum effects in the UEG.

In studies of the SSF for classic OCP systems, the mean spherical approximation yields results that are in excellent agreement with Monte Carlo simulations~\cite{Singh1983} and the experimental data for liquid alkali metals~\cite{Singh1983PRA}. Within this theoretical framework, the only undetermined parameter is the packing fraction $\eta$, which determines the effective diameter of the charged particles~\cite{Singh1983PRA}. It can be computed using the following fitting formula
\begin{equation}\label{diameter}
    \eta_\mathrm{fit}(t) = \frac{t^2 + c_1 t^3 + c_2 t^4 + c_3 t^5}{8 + 8 b_1 t + b_2 t^2 + b_3 t^3 + b_4 t^4 + c_3 t^5}
\end{equation}
with $b_1 = c_1 + 3^{3/2}, b_3 = c_1 + 108^{1/6} c_2 + 32^{1/3} c_3, b_4 = c_2 + 108^{1/6} c_3$. The fitting parameters are $c_1 = 9962.883, c_2 = -14619.16, c_3 = 11211.48, b_2 = 6344.277$. Here $t = t(g) = g^{3/2} / (1 + g^{4/3})$. For classical OCP systems, $g = \Gamma$ is adopted. A comparative analysis of the UEG SSF reveals that, over a broad range of temperatures and densities, the value of $\alpha_2$ can be derived from the expression~\eqref{diameter} using $\alpha_2 = 2.5 \, \eta_\mathrm{fit}^{1/3}$ with $g=\alpha_0$, while the value of $\alpha_1$ is still determined by solving Eq.~\eqref{g0func}.

Fig.~\ref{fig:sfg2fix} presents a comparison among the results computed using the numeric strategy described in Sec.~\ref{sec:theory:ssf}, those obtained from Eq.~\eqref{diameter}, and the reference PIMC data. For plasma conditions corresponding to $\rmcapsub{r}{S} = 10$ and $\theta = 4$, both methods produce results for the SSF that agree excellently with the PIMC benchmarks. As the coupling strength increases, discrepancies between the two approaches become apparent, particularly in the peak region of the SSF. Under strong coupling conditions $(\rmcapsub{r}{S} = 100)$, although the model proposed in this work does not fully reproduce the peak compared to the PIMC results, the method based on Eq.~\eqref{diameter} shows better agreement with PIMC data within the range $q \leq 4$. Moreover, it more accurately captures the peak position and begins to exhibit oscillatory features. These findings suggest that refining the constraints used to determine $\alpha_1$ and $\alpha_2$ may further enhance the performance of the current SSF model.

\subsection{\label{sec:results:dielectric}Static dielectric function}
\begin{figure}[t]
\includegraphics[width=0.45 \textwidth]{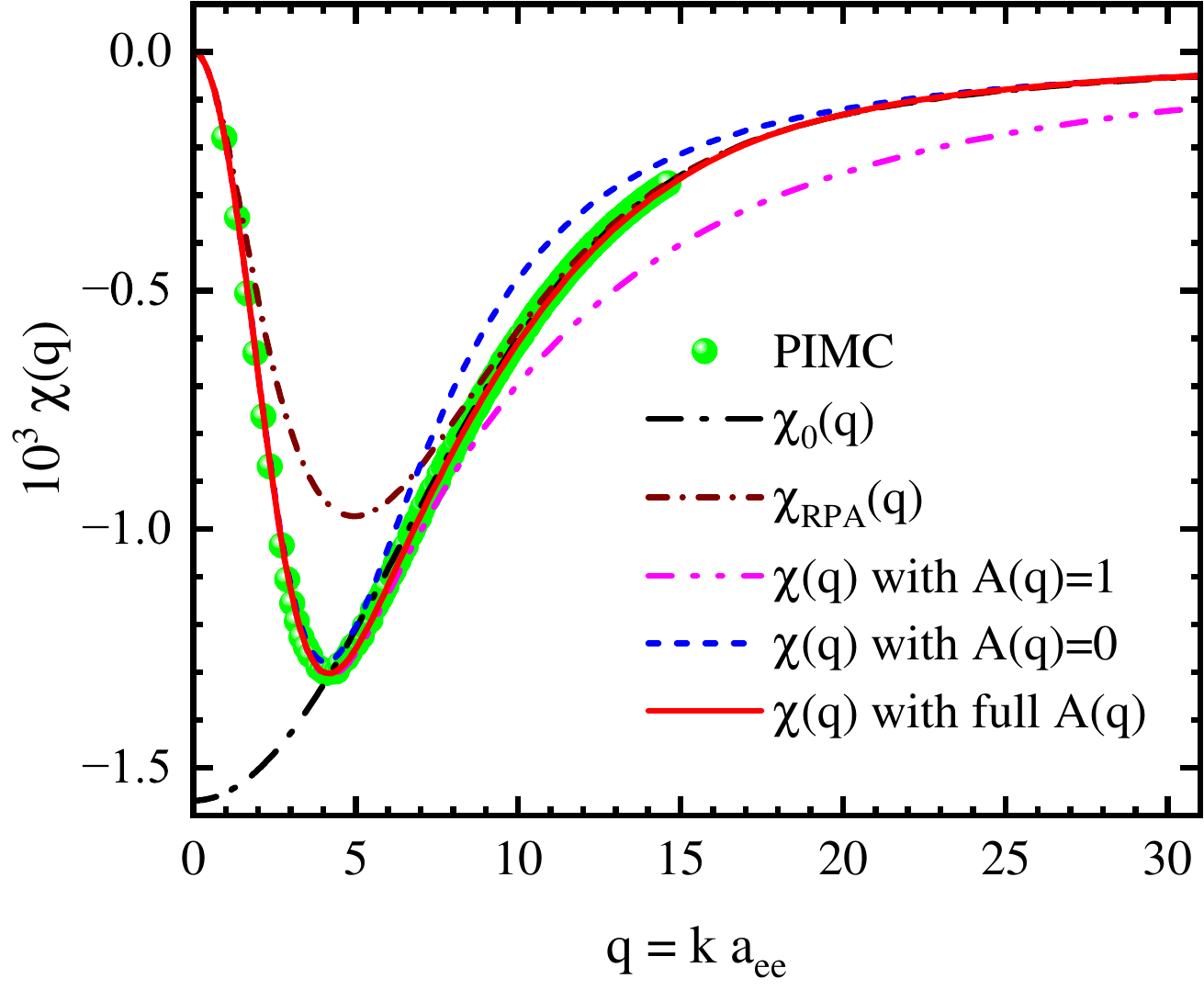}
\caption{\label{fig3_chiq_kq}The static response function $\chi(q)$ for $\rmcapsub{r}{S} = 20$ and $\theta = 4$ predicted by different approaches. The square symbols represent the PIMC data reported in Ref.~\cite{Dornheim2020}. The magenta, blue and red curves represent the predictions calculated using $A(q)=1$ (magenta), $A(q)=0$ (blue) and $A(q)$~\eqref{aqfunc} (red), respectively. For comparison, the results for Lindhard response function $\chi_0(q)$ (black) and the RPA response function $\rmcapsub{\chi}{RPA}(q)$ (brown) are also displayed.}
\end{figure}

In this section, we discuss the static response properties of UEG. Fig.~\ref{fig3_chiq_kq} compares the static density response functions obtained from the present model under different choice of parameter function $A(q)$, with several commonly used theoretical approximations and reference PIMC results. As illustrated, the RPA static density response function $\rmcapsub{\chi}{RPA}(q)$ fails to accurately capture the response within the range $2<q<10$, primarily due to its neglect of Coulomb coupling effects. On the other hand, the Lindhard response function $\chi_0(q)$ of the ideal electron gas does not account for the collective response behavior, rendering it inaccurate in the long-wavelength limit. Unlike other models, $\chi_0(q)$ approaches a finite value in the long-wavelength limit as $q \rightarrow 0$ 
\begin{equation}
    \chi_0(q = 0) = \frac{\theta^{1/2}}{2 \pi^2 \gamma_0\, \rmcapsub{r}{S}} F_{-1/2}(\eta).
\end{equation}

\begin{figure}[t]
\includegraphics[width=0.45 \textwidth]{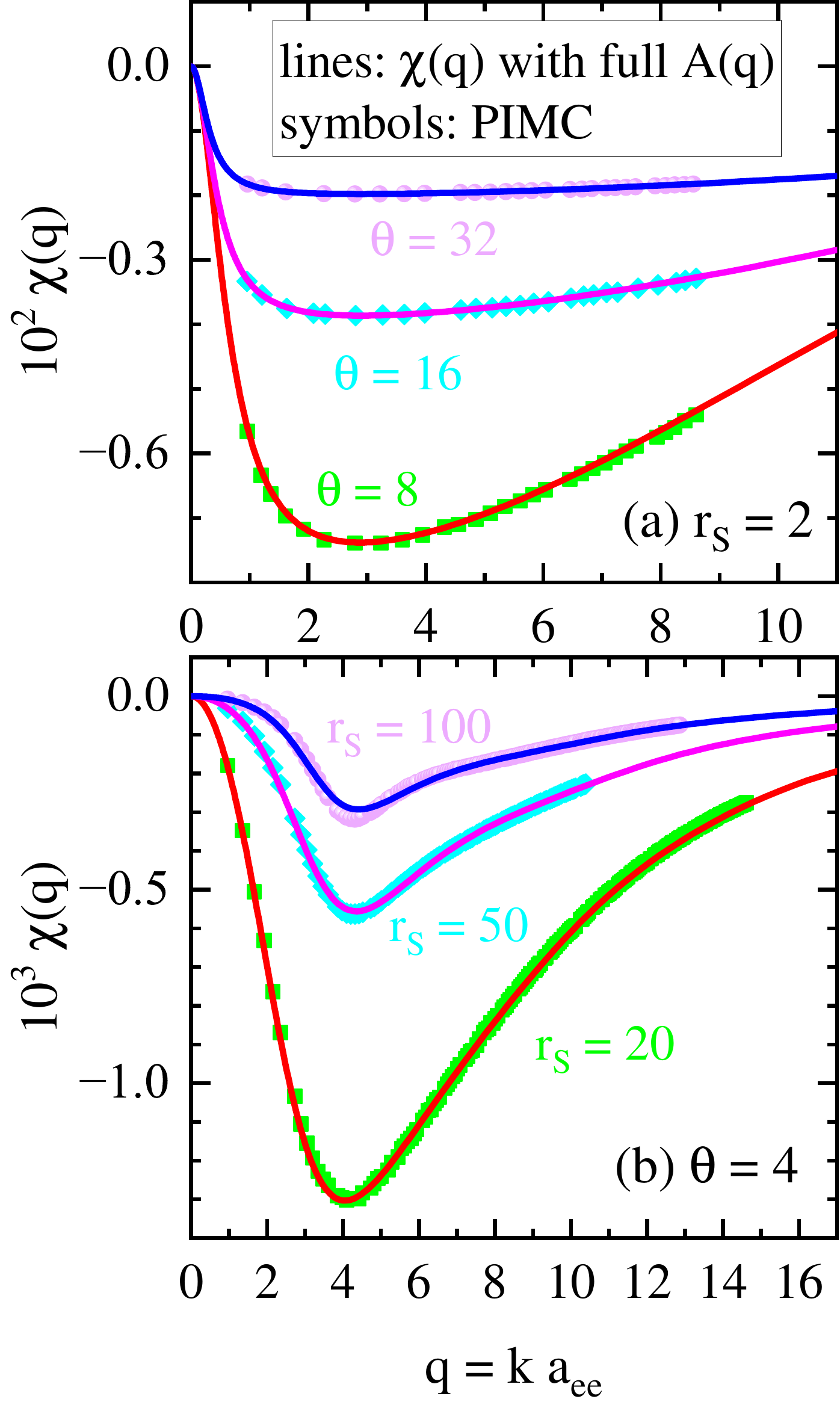}
\caption{\label{fig4_chiq_rst}The static response function $\chi(q)$ for a fixed density $\rmcapsub{r}{S} = 2$ at three degeneracy parameters (upper panel (a)) and for a fixed degeneracy $\theta = 4$ at three electron densities (lower panel (b)). The PIMC data are taken from Ref.~\cite{Dornheim2022} for $\rmcapsub{r}{S} = 2$ in the panel (a) and from Ref.~\cite{Dornheim2020} for $\theta = 4$ in the panel (b). The lines represent the results calculated using the model proposed in this work.}
\end{figure}

In Sec.~\ref{sec:theory:ueg}, various models for the parameter function $A(q)$ were discussed. The corresponding results with different choices of $A(q)$ for $\rmcapsub{r}{S} = 20$ and $\theta = 4$ are also presented in Fig.~\ref{fig3_chiq_kq}. It can be seen that the model with $A(q) = 1$ accurately describes the static response behavior within the range $0<q<6$. However, as $q$ increases beyond this range, deviations from PIMC results become increasingly noticeable. In contrast, the model with $A(q)=0$ begins to deviate from the PIMC data near the minimum of the response function (around $q \approx 4$). However, compared to the case of $A(q) = 1$, the model with $A(q) = 0$ exhibits better overall agreement with the PIMC trends and converges more rapidly toward RPA behavior as $q$ increases. These results suggest that $A(q)$ decays from a finite value to zero as $q$ increases. Since the long-wavelength response is insensitive to the finite value of $A(q)$ at small $q$, and since the model with $A(q) = 1$ performs well for $q < 6$, we propose the expression given in Eq.~\eqref{aqfunc} to represent the $q$-dependence of $A(q)$.

To further validate the effectiveness of Eq.~\eqref{aqfunc}, we calculate the static density response function $\chi(q)$ of UEG under various temperature and density conditions, and compare the results with corresponding PIMC simulations~\cite{Dornheim2022,Dornheim2020}, as shown in Fig.~\ref{fig4_chiq_rst}. The results obtained from Eq.~\eqref{aqfunc} show excellent agreement with the PIMC data in all conditions considered. Even in extremely low-density regimes (e.g., $\rmcapsub{r}{S} = 50$ and $100$), the predictions based on Eq.~\eqref{aqfunc} remain consistent with the overall trends observed in the simulations. The discrepancies near the minimum are primarily due to an inadequate description of the coupling peak in the SSF as shown in Fig.~\ref{fig:sfg1g2}, and may also be affected by inaccuracies in the parameter function $A(q)$ within this region. Addressing these issues will require further refinement of both the SSF model and a more precise representation of the parameter function $A(q)$. Furthermore, due to the absence of PIMC data at higher $q$ values, it is not yet clear whether the form of $A(q)$ should be adjusted to account for the effects of temperature and density.

\subsection{\label{sec:results:stopping}Low-velocity stopping power and friction coefficient}

Finally, in this subsection, we turn to the electron–ion friction coefficient. In Langevin dynamics simulations, the friction coefficient based on the Rayleigh model is widely adopted, which takes the following form~\cite{Dai2010,Fu2018,Hou2021}
\begin{equation}\label{rayleigh}
    \rmcapsub{\xi}{R} = m_\mathrm{i} \rmcapsub{\gamma}{R}  = 2 \pi Z^{2/3} \xi_0 .
\end{equation}
Furthermore, Zwicknagel~\cite{Zwicknagel2002} investigated the energy loss of charged particles in an electron gas across various coupling regimes and provided a fitted expression for the friction coefficient based on low-velocity stopping power. Adapted to the notation used in this work, their expression takes the form
\begin{equation}\label{zwicknagel}
    \rmcapsub{\xi}{Z} = \xi_0 \frac{3 Z^2 \Gamma^2 }{8} \squarebra{\ln\bks{1 + \frac{0.14}{x^2}} + \frac{1.8}{1 + 0.4\, x^{1.3}}} 
\end{equation}
with $x = Z\, \Gamma^{3/2}$.

\begin{figure}[t]
\includegraphics[width=0.48 \textwidth]{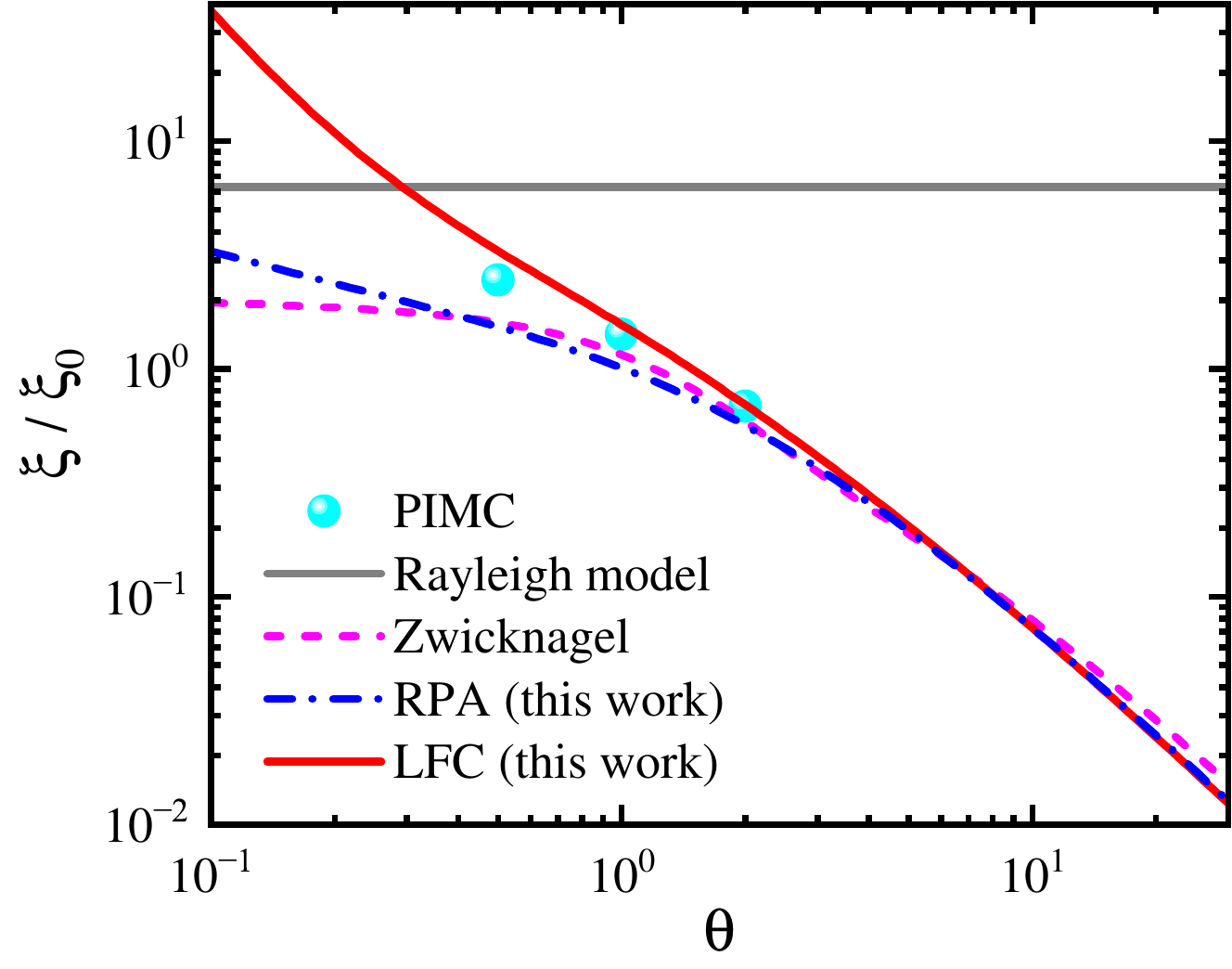}
\caption{\label{fig5_friction}The reduced friction coefficient at the electron density $\rmcapsub{r}{S} = 4$ with varying degeneracy $\theta$. The PIMC data are taken from Ref.~\cite{Moldabekov2020}. The gray and magenta lines present the results from the Rayleigh model~\eqref{rayleigh} and the results based on the fit of Zwicknagel~\eqref{zwicknagel}. The predictions denoted by RPA (blue) and LFC (red) are calculated without and with the static local field correction from Eq.~\eqref{frictionlfcFfunc}, respectively.}
\end{figure}
In Fig.~\ref{fig5_friction}, we present a comparison of the reduced electron-ion friction coefficient $\xi/\xi_0$ predicted by different models at $\rmcapsub{r}{S} = 4$, examining its behavior as a function of the electron degeneracy parameter $\theta$. The Rayleigh model~\eqref{rayleigh} yields a reduced friction coefficient that remains constant with varying degeneracy, while all other models show an increase as $\theta$ decreases (i.e., as degeneracy effects strengthen). Although the absolute values differ, both the RPA model (Eq.~\eqref{frictionlfcFfunc} with the local field correction $G(q) = 0$, see also Ref.~\cite{Arista1985JPC}) and Zwicknagel's fit (Eq.~\eqref{zwicknagel}) exhibit a rising trend with decreasing $\theta$. In contrast, the LFC-based model agrees more closely with PIMC results at $\theta = 2$ and $\theta = 1$. However, at $\theta=0.5\  (\Gamma = 4.3)$, it predicts a value of $0.206$, exceeding the PIMC benchmark of $0.154$. As $\theta$ decreases further, the LFC model shows a faster increase in the friction coefficient, which may stem from inaccuracies in the SSF under strong coupling conditions.

In Langevin molecular dynamics, the friction coefficient $\gamma = \xi / m_\mathrm{i}$ quantifies the influence of the damping of medium on particle motion. It underscores the role of fluctuation–dissipation mechanisms in shaping plasma transport properties, such as ion diffusion, viscosity, and thermal conductivity. By governing the temporal and spatial scales of energy dissipation, the friction coefficient critically influences the dynamical structure factor, sound waves of the nonideal ions, and the non-equilibrium behavior of the system~\cite{Moldabekov2020,Hou2021,Mabey2017,Kaehlert2019}. The comparison presented in Fig.~\ref{fig5_friction} indicates the need for further investigation into the friction coefficient.

\section{\label{sec:conclusions}Conclusions}
Based on the comprehensive analysis presented in this work, we have developed and systematically validated a robust analytical model for the SSF and static density response function of the UEG, applicable across a broad range of temperatures and densities. By combining physically motivated analytical form with constraints from highly accurate PIMC simulations, the proposed model accurately reproduces key characteristics of the SSF within the parameter region $1 \le \rmcapsub{r}{S} \le 10$ and $\theta \ge 1$, including the position and height of the correlation peak. The model also demonstrates good agreement with PIMC benchmarks under stronger coupling conditions, although discrepancies emerge at very low temperatures and high $\rmcapsub{r}{S}$ values, particularly in the oscillatory behavior of the SSF.

The parameter function $A(q)$, introduced to ensure correct asymptotic behavior of the static density response function, plays a critical role in bridging the gap between the random phase approximation and strongly correlated regimes. Our findings demonstrate that a $q$-dependent form of $A(q)$, as specified in Eq.~\eqref{aqfunc}, significantly enhances agreement with PIMC data across varying coupling strengths and degeneracy conditions. Additionally, the electron–ion friction coefficient $\xi$, derived from the low-velocity limit of the stopping power, exhibits reasonable consistency with PIMC results at moderate degeneracy. Some overestimation occurs under strong coupling, indicating a need for further refinement of the SSF model in these regimes.

Despite these advances, several challenges remain. The current SSF model does not fully capture the oscillatory behavior observed in PIMC simulations at low temperatures and densities, highlighting the necessity of more accurate treatments of the interplay between Coulomb coupling and quantum degeneracy. The analytical framework presented here offers a foundation for future improvements. Subsequent work will focus on developing more advanced theoretical approaches that adhere to additional sum rules and asymptotic constraints. Such enhancements will improve the description of both static and dynamic response properties, enabling more reliable simulations of energy deposition, stopping power, and non-equilibrium dynamics in warm dense matter and other strongly correlated systems.

\begin{acknowledgments}
This work was supported by the Science Challenge Project under Grants No. TZ2025013 and the National Natural Science Foundation of China under Grants Nos. 12474277, U2430208, 12274384, and 11974424.
\end{acknowledgments}

\section*{DATA AVAILABILITY}
The data that supports the findings of this study are available from the corresponding author upon reasonable request.

\nocite{*}

\bibliography{apssamp}

@PREAMBLE{
 "\providecommand{\noopsort}[1]{}" 
 # "\providecommand{\singleletter}[1]{#1}%" 
}

@BOOK{Nozieres1999,
   author       = {Ph. Nozi\`eres and D. Pines},
   year         = 1999,
   title        = {Theory of Quantum Liquids},
   publisher    = {Avalon Publishing}
}

@BOOK{Giuliani2008,
   author       = {G. Giuliani and G. Vignale},
   year         = 2008,
   title        = {Quantum Theory of the Electron
Liquid},
   publisher    = {Cambridge University Press}
}

@BOOK{Isihara1998,
   author       = {Akira Isihara},
   year         = 1998,
   title        = {Electron Liquids},
   publisher    = {Springer-Verlag Berlin Heidelberg}
}

@article{Kim2022,
    author = {Sunghun Kim and Joonho Bang and Chan Young Lim and Seung Yong Lee and Jounghoon Hyun and Gyubin Lee and Yeonghoon Lee and Jonathan D. Denlinger and Soonsang Huh and Changyoung Kim and Sang Yong Song and Jungpil Seo and Dinesh Thapa and Seong-Gon Kim and Young Hee Lee and Yeongkwan Kim and Sung Wng Kim},
    title = {Quantum electron liquid and its possible phase transition},
    journal = {Nat. Mater.},
    volume = {21},
    pages = {1269–1274},
    year = {2022},
    doi = {10.1038/s41563-022-01353-8},
    url = {https://doi.org/10.1038/s41563-022-01353-8}
}

@article{Dornheim2025,
    author = {Tobias Dornheim and Tilo D\"oppner and Panagiotis Tolias and Maximilian P. B\"ohme and Luke B. Fletcher and Thomas Gawne and Frank R. Graziani and Dominik Kraus and Michael J. MacDonald and Zhandos A. Moldabekov and Sebastian Schwalbe and Dirk O. Gericke and Jan Vorberger},
    title = {Unraveling electronic correlations in warm dense quantum plasmas},
    journal = {Nat. Commun.},
    volume = {16},
    pages = {5103},
    year = {2025},
    doi = {10.1038/s41467-025-60278-3},
    url = {https://doi.org/10.1038/s41467-025-60278-3}
}

@article{Glenzer2004,
  title = {X-ray Thomson scattering in high energy density plasmas},
  author = {Glenzer, Siegfried H. and Redmer, Ronald},
  journal = {Rev. Mod. Phys.},
  volume = {81},
  issue = {4},
  pages = {1625--1663},
  numpages = {0},
  year = {2009},
  month = {Dec},
  publisher = {American Physical Society},
  doi = {10.1103/RevModPhys.81.1625},
  url = {https://link.aps.org/doi/10.1103/RevModPhys.81.1625}
}

@article{Brown2013,
  title = {Path-Integral Monte Carlo Simulation of the Warm Dense Homogeneous Electron Gas},
  author = {Brown, Ethan W. and Clark, Bryan K. and DuBois, Jonathan L. and Ceperley, David M.},
  journal = {Phys. Rev. Lett.},
  volume = {110},
  issue = {14},
  pages = {146405},
  numpages = {5},
  year = {2013},
  month = {Apr},
  publisher = {American Physical Society},
  doi = {10.1103/PhysRevLett.110.146405},
  url = {https://link.aps.org/doi/10.1103/PhysRevLett.110.146405}
}

@article{Dornheim2018,
title = {The uniform electron gas at warm dense matter conditions},
journal = {Phys. Rep.},
volume = {744},
pages = {1-86},
year = {2018},
issn = {0370-1573},
doi = {https://doi.org/10.1016/j.physrep.2018.04.001},
url = {https://www.sciencedirect.com/science/article/pii/S0370157318300516},
author = {Dornheim, Tobias and Groth, Simon and Bonitz, Michael}
}

@article{Zaghoo2019,
  title = {Breakdown of Fermi Degeneracy in the Simplest Liquid Metal},
  author = {Zaghoo, M. and Boehly, T. R. and Rygg, J. R. and Celliers, P. M. and Hu, S. X. and Collins, G. W.},
  journal = {Phys. Rev. Lett.},
  volume = {122},
  issue = {8},
  pages = {085001},
  numpages = {6},
  year = {2019},
  month = {Feb},
  publisher = {American Physical Society},
  doi = {10.1103/PhysRevLett.122.085001},
  url = {https://link.aps.org/doi/10.1103/PhysRevLett.122.085001}
}

@article{Kodanova2025,
    author = {Kodanova, S. K. and Ramazanov, T. S. and Issanova, M. K.},
    title = {Impact of local field correction on transport and dynamic properties of warm dense matter},
    journal = {Matter Radiat. Extrem.},
    volume = {10},
    number = {3},
    pages = {037601},
    year = {2025},
    month = {02},
    doi = {10.1063/5.0243102},
    url = {https://doi.org/10.1063/5.0243102}
}

@article{Ding2018,
  title = {Ab Initio Studies on the Stopping Power of Warm Dense Matter with Time-Dependent Orbital-Free Density Functional Theory},
  author = {Ding, Y. H. and White, A. J. and Hu, S. X. and Certik, O. and Collins, L. A.},
  journal = {Phys. Rev. Lett.},
  volume = {121},
  issue = {14},
  pages = {145001},
  numpages = {6},
  year = {2018},
  month = {Oct},
  publisher = {American Physical Society},
  doi = {10.1103/PhysRevLett.121.145001},
  url = {https://link.aps.org/doi/10.1103/PhysRevLett.121.145001}
}

@article{Hentschel2023,
    author = {Hentschel, Thomas W. and Kononov, Alina and Olmstead, Alexandra and Cangi, Attila and Baczewski, Andrew D. and Hansen, Stephanie B.},
    title = {Improving dynamic collision frequencies: Impacts on dynamic structure factors and stopping powers in warm dense matter},
    journal = {Phys. Plasmas},
    volume = {30},
    number = {6},
    pages = {062703},
    year = {2023},
    month = {06},
    doi = {10.1063/5.0143738},
    url = {https://doi.org/10.1063/5.0143738}
}

@article{He2018,
    author = {He, Bin and Wang, Zhi-Gang and Wang, Jian-Guo},
    title = {Slowing down of alpha particles in ICF DT plasmas},
    journal = {Phys. Plasmas},
    volume = {25},
    number = {1},
    pages = {012704},
    year = {2018},
    month = {01},
    doi = {10.1063/1.5004213},
    url = {https://doi.org/10.1063/1.5004213}
}

@article{SM2016,
	title = {Ionic transport in high-energy-density matter},
	author = {Stanton, Liam G. and Murillo, Michael S.},
	journal = {Phys. Rev. E},
	volume = {93},
	issue = {4},
	pages = {043203},
	numpages = {23},
	year = {2016},
	month = {Apr},
	publisher = {American Physical Society},
	doi = {10.1103/PhysRevE.93.043203},
	url = {https://link.aps.org/doi/10.1103/PhysRevE.93.043203}
}

@article{Hu2024,
    author = {Hu, S. X. and Nichols, K. A. and Shaffer, N. R. and Arnold, B. and White, A. J. and Collins, L. A. and Karasiev, V. V. and Zhang, S. and Goncharov, V. N. and Shah, R. C. and Mihaylov, D. I. and Jiang, S. and Ping, Y.},
    title = {A review on charged-particle transport modeling for laser direct-drive fusion},
    journal = {Phys. Plasmas},
    volume = {31},
    number = {4},
    pages = {040501},
    year = {2024},
    month = {04},
    doi = {10.1063/5.0197969},
    url = {https://doi.org/10.1063/5.0197969}
}

@article{Hayes2020,
    author = {A. C. Hayes and M. E. Gooden and E. Henry and Gerard Jungman and J. B. Wilhelmy and R. S. Rundberg and C. Yeamans and G. Kyrala and C. Cerjan and D. L. Danielson and Jérôme Daligault and C. Wilburn and P. Volegov and C. Wilde and S. Batha and T. Bredeweg and J. L. Kline and G. P. Grim and E. P. Hartouni and D. Shaughnessy and C. Velsko and W. S. Cassata and K. Moody and L. F. Berzak Hopkins and D. Hinkel and T. D\"oppner and S. Le Pape and F. Graziani and D. A. Callahan and O. A. Hurricane  and D. Schneider},
    title = {Plasma stopping-power measurements reveal transition from non-degenerate to degenerate plasmas},
    journal = {Nat. Phys.},
    volume = {16},
    pages = {432–437},
    year = {2020},
    doi = {10.1038/s41567-020-0790-3},
    url = {https://doi.org/10.1038/s41567-020-0790-3}
}

@article{Kraus2017,
    author = {D. Kraus and J. Vorberger and A. Pak and N. J. Hartley and L. B. Fletcher and S. Frydrych and E. Galtier and E. J. Gamboa and D. O. Gericke and S. H. Glenzer and E. Granados and M. J. MacDonald and A. J. MacKinnon and E. E. McBride and I. Nam and P. Neumayer and M. Roth and A. M. Saunders and A. K. Schuster and P. Sun and T. van Driel and T. D\"oppner and R. W. Falcone },
    title = {Formation of diamonds in laser-compressed hydrocarbons at planetary interior conditions},
    journal = {Nat. Astron.},
    volume = {1},
    pages = {606–611},
    year = {2017},
    doi = {10.1038/s41550-017-0219-9},
    url = {https://doi.org/10.1038/s41550-017-0219-9}
}

@article{Vorberger2007,
  title = {Hydrogen-helium mixtures in the interiors of giant planets},
  author = {Vorberger, J. and Tamblyn, I. and Militzer, B. and Bonev, S. A.},
  journal = {Phys. Rev. B},
  volume = {75},
  issue = {2},
  pages = {024206},
  numpages = {11},
  year = {2007},
  month = {Jan},
  publisher = {American Physical Society},
  doi = {10.1103/PhysRevB.75.024206},
  url = {https://link.aps.org/doi/10.1103/PhysRevB.75.024206}
}

@article{Chabrier1993,
    author = {Chabrier, Gilles},
    title = {Quantum Effects in Dense Coulombic Matter: Application to the Cooling of White Dwarfs},
    journal = {Astrophys. J.},
    volume = {414},
    pages = {695},
    year = {1993},
    doi = {10.1086/173115},
    url = {https://doi.org/10.1086/173115}
}

@article{Ichimaru1987,
title = {Statistical physics of dense plasmas: Thermodynamics, transport coefficients and dynamic correlations},
journal = {Phys. Rep.},
volume = {149},
number = {2},
pages = {91-205},
year = {1987},
issn = {0370-1573},
doi = {https://doi.org/10.1016/0370-1573(87)90125-6},
url = {https://www.sciencedirect.com/science/article/pii/0370157387901256},
author = {Setsuo Ichimaru and Hiroshi Iyetomi and Shigenori Tanaka}
}

@article{Lin2016,
  title = {Ionization-potential depression and dynamical structure factor in dense plasmas},
  author = {Lin, Chengliang and R\"opke, Gerd and Kraeft, Wolf-Dietrich and Reinholz, Heidi},
  journal = {Phys. Rev. E},
  volume = {96},
  issue = {1},
  pages = {013202},
  numpages = {8},
  year = {2017},
  month = {Jul},
  publisher = {American Physical Society},
  doi = {10.1103/PhysRevE.96.013202},
  url = {https://link.aps.org/doi/10.1103/PhysRevE.96.013202}
}

@article{Lin2019,
    author = {Lin, Chengliang},
    title = {Quantum statistical approach for ionization potential depression in multi-component dense plasmas},
    journal = {Phys. Plasmas},
    volume = {26},
    number = {12},
    pages = {122707},
    year = {2019},
    month = {12},
    doi = {10.1063/1.5124544},
    url = {https://doi.org/10.1063/1.5124544}
}

@article{Zan2021,
  title = {Local field correction to ionization potential depression of ions in warm or hot dense matter},
  author = {Zan, Xiaolei and Lin, Chengliang and Hou, Yong and Yuan, Jianmin},
  journal = {Phys. Rev. E},
  volume = {104},
  issue = {2},
  pages = {025203},
  numpages = {10},
  year = {2021},
  month = {Aug},
  publisher = {American Physical Society},
  doi = {10.1103/PhysRevE.104.025203},
  url = {https://link.aps.org/doi/10.1103/PhysRevE.104.025203}
}

@article{Utsumi1980,
  title = {Dielectric formulation of strongly coupled electron liquids at metallic densities. II. Exchange effects and static properties},
  author = {Utsumi, Kenichi and Ichimaru, Setsuo},
  journal = {Phys. Rev. B},
  volume = {22},
  issue = {11},
  pages = {5203--5212},
  numpages = {0},
  year = {1980},
  month = {Dec},
  publisher = {American Physical Society},
  doi = {10.1103/PhysRevB.22.5203},
  url = {https://link.aps.org/doi/10.1103/PhysRevB.22.5203}
}

@article{Karasiev2016,
  title = {Importance of finite-temperature exchange correlation for warm dense matter calculations},
  author = {Karasiev, Valentin V. and Calder\'{\i}n, L\'azaro and Trickey, S. B.},
  journal = {Phys. Rev. E},
  volume = {93},
  issue = {6},
  pages = {063207},
  numpages = {12},
  year = {2016},
  month = {Jun},
  publisher = {American Physical Society},
  doi = {10.1103/PhysRevE.93.063207},
  url = {https://link.aps.org/doi/10.1103/PhysRevE.93.063207}
}

@article{Groth2017,
  title = {Ab initio Exchange-Correlation Free Energy of the Uniform Electron Gas at Warm Dense Matter Conditions},
  author = {Groth, Simon and Dornheim, Tobias and Sjostrom, Travis and Malone, Fionn D. and Foulkes, W. M. C. and Bonitz, Michael},
  journal = {Phys. Rev. Lett.},
  volume = {119},
  issue = {13},
  pages = {135001},
  numpages = {7},
  year = {2017},
  month = {Sep},
  publisher = {American Physical Society},
  doi = {10.1103/PhysRevLett.119.135001},
  url = {https://link.aps.org/doi/10.1103/PhysRevLett.119.135001}
}

@article{Hilleke2025,
  title = {Fully thermal meta-GGA exchange correlation free-energy density functional},
  author = {Hilleke, K. P. and Karasiev, V. V. and Trickey, S. B. and Goshadze, R. M. N. and Hu, S. X.},
  journal = {Phys. Rev. Mater.},
  volume = {9},
  issue = {5},
  pages = {L050801},
  numpages = {8},
  year = {2025},
  month = {May},
  publisher = {American Physical Society},
  doi = {10.1103/PhysRevMaterials.9.L050801},
  url = {https://link.aps.org/doi/10.1103/PhysRevMaterials.9.L050801}
}

@article{Dharma2000,
  title = {Simple Classical Mapping of the Spin-Polarized Quantum Electron Gas: Distribution Functions and Local-Field Corrections},
  author = {Dharma-wardana, M. W. C. and Perrot, F.},
  journal = {Phys. Rev. Lett.},
  volume = {84},
  issue = {5},
  pages = {959--962},
  numpages = {0},
  year = {2000},
  month = {Jan},
  publisher = {American Physical Society},
  doi = {10.1103/PhysRevLett.84.959},
  url = {https://link.aps.org/doi/10.1103/PhysRevLett.84.959}
}

@BOOK{Kremp2005,
   author       = {Kremp, D. and Schlanges, M. and Kraeft, W.-D.},
   year         = 2005,
   title        = {Quantum Statistics of Nonideal Plasmas},
   publisher    = {Springer-Verlag Berlin Heidelberg}
}

@article{Ornstein1914,
  title = {Accidental deviations of density and opalescence at the critical point of a single substance},
  author = {Ornstein, L. S. and Zernike, F.},
  journal = {Proc. Acad. Sci. Amsterdam},
  volume = {17},
  pages = {793},
  year = {1914}
}

@article{Baus1980,
title = {Statistical mechanics of simple coulomb systems},
journal = {Phys. Rep.},
volume = {59},
number = {1},
pages = {1-94},
year = {1980},
issn = {0370-1573},
doi = {https://doi.org/10.1016/0370-1573(80)90022-8},
url = {https://www.sciencedirect.com/science/article/pii/0370157380900228},
author = {Baus, Marc and Hansen, Jean-Pierre}
}

@BOOK{Hansen2006,
   author       = {Hansen, Jean-Pierre and McDonald, I.R.},
   year         = 2006,
   title        = {Theory of Simple Liquids},
   publisher    = {Academic Press}
}

@article{Fortmann2010,
  title = {Influence of local-field corrections on Thomson scattering in collision-dominated two-component plasmas},
  author = {Fortmann, Carsten and Wierling, August and R\"opke, Gerd},
  journal = {Phys. Rev. E},
  volume = {81},
  issue = {2},
  pages = {026405},
  numpages = {11},
  year = {2010},
  month = {Feb},
  publisher = {American Physical Society},
  doi = {10.1103/PhysRevE.81.026405},
  url = {https://link.aps.org/doi/10.1103/PhysRevE.81.026405},
  note={and references thererin.}
}

@article{Dornheim2021,
  title = {Density response of the warm dense electron gas beyond linear response theory: Excitation of harmonics},
  author = {Dornheim, Tobias and B\"ohme, Maximilian and Moldabekov, Zhandos A. and Vorberger, Jan and Bonitz, Michael},
  journal = {Phys. Rev. Res.},
  volume = {3},
  issue = {3},
  pages = {033231},
  numpages = {23},
  year = {2021},
  month = {Sep},
  publisher = {American Physical Society},
  doi = {10.1103/PhysRevResearch.3.033231},
  url = {https://link.aps.org/doi/10.1103/PhysRevResearch.3.033231}
}

@article{Kaehlert2024,
    author = {K\"ahlert, Hanno},
    title = {Dynamic local field correction of the one-component plasma},
    journal = {Phy. Plasmas},
    volume = {31},
    number = {9},
    pages = {092109},
    year = {2024},
    month = {09},
    doi = {10.1063/5.0229805},
    url = {https://doi.org/10.1063/5.0229805}
}

@article{Atwal2002,
  title = {Relaxation of an electron system: Conserving approximation},
  author = {Atwal, G. S. and Ashcroft, N. W.},
  journal = {Phys. Rev. B},
  volume = {65},
  issue = {11},
  pages = {115109},
  numpages = {14},
  year = {2002},
  month = {Feb},
  publisher = {American Physical Society},
  doi = {10.1103/PhysRevB.65.115109},
  url = {https://link.aps.org/doi/10.1103/PhysRevB.65.115109}
}

@article{Chuna2025,
  title = {Conservative dielectric functions and electrical conductivities from the multicomponent Bhatnagar-Gross-Krook equation},
  author = {Chuna, Thomas and Murillo, Michael S.},
  journal = {Phys. Rev. E},
  volume = {111},
  issue = {3},
  pages = {035206},
  numpages = {16},
  year = {2025},
  month = {Mar},
  publisher = {American Physical Society},
  doi = {10.1103/PhysRevE.111.035206},
  url = {https://link.aps.org/doi/10.1103/PhysRevE.111.035206}
}

@article{Groth2019,
  title = {Ab initio path integral Monte Carlo approach to the static and dynamic density response of the uniform electron gas},
  author = {Groth, S. and Dornheim, T. and Vorberger, J.},
  journal = {Phys. Rev. B},
  volume = {99},
  issue = {23},
  pages = {235122},
  numpages = {20},
  year = {2019},
  month = {Jun},
  publisher = {American Physical Society},
  doi = {10.1103/PhysRevB.99.235122},
  url = {https://link.aps.org/doi/10.1103/PhysRevB.99.235122}
}

@article{Filinov2020,
  title = {Uniform electron gas at finite temperature by fermionic-path-integral Monte Carlo simulations},
  author = {Filinov, V. S. and Larkin, A. S. and Levashov, P. R.},
  journal = {Phys. Rev. E},
  volume = {102},
  issue = {3},
  pages = {033203},
  numpages = {11},
  year = {2020},
  month = {Sep},
  publisher = {American Physical Society},
  doi = {10.1103/PhysRevE.102.033203},
  url = {https://link.aps.org/doi/10.1103/PhysRevE.102.033203}
}

@article{Larkin2021,
    author = {Larkin, A. S. and Filinov, V. S. and Levashov, P. R.},
    title = {Single-momentum path integral Monte Carlo simulations of uniform electron gas in warm dense matter regime},
    journal = {Phys. Plasmas},
    volume = {28},
    number = {12},
    pages = {122712},
    year = {2021},
    month = {12},
    doi = {10.1063/5.0072354},
    url = {https://doi.org/10.1063/5.0072354}
}

@article{Dornheim2020,
  title = {Strongly coupled electron liquid: Ab initio path integral Monte Carlo simulations and dielectric theories},
  author = {Dornheim, Tobias and Sjostrom, Travis and Tanaka, Shigenori and Vorberger, Jan},
  journal = {Phys. Rev. B},
  volume = {101},
  issue = {4},
  pages = {045129},
  numpages = {18},
  year = {2020},
  month = {Jan},
  publisher = {American Physical Society},
  doi = {10.1103/PhysRevB.101.045129},
  url = {https://link.aps.org/doi/10.1103/PhysRevB.101.045129}
}

@article{Tanaka2016,
    author = {Tanaka, Shigenori},
    title = {Correlational and thermodynamic properties of finite-temperature electron liquids in the hypernetted-chain approximation},
    journal = {J. Chem. Phys.},
    volume = {145},
    number = {21},
    pages = {214104},
    year = {2016},
    month = {12},
    doi = {10.1063/1.4969071},
    url = {https://doi.org/10.1063/1.4969071}
}

@article{Dornheim2017,
author = {Dornheim, Tobias and Groth, Simon and Bonitz, Michael},
title = {Ab initio results for the static structure factor of the warm dense electron gas},
journal = {Contrib. Plasma Phys.},
volume = {57},
number = {10},
pages = {468-478},
doi = {https://doi.org/10.1002/ctpp.201700096},
url = {https://onlinelibrary.wiley.com/doi/abs/10.1002/ctpp.201700096},
year = {2017}
}

@article{Dornheim2020letter,
  title = {Effective Static Approximation: A Fast and Reliable Tool for Warm-Dense Matter Theory},
  author = {Dornheim, Tobias and Cangi, Attila and Ramakrishna, Kushal and B\"ohme, Maximilian and Tanaka, Shigenori and Vorberger, Jan},
  journal = {Phys. Rev. Lett.},
  volume = {125},
  issue = {23},
  pages = {235001},
  numpages = {8},
  year = {2020},
  month = {Dec},
  publisher = {American Physical Society},
  doi = {10.1103/PhysRevLett.125.235001},
  url = {https://link.aps.org/doi/10.1103/PhysRevLett.125.235001}
}

@article{Zwicknagel2002,
title = {Nonlinear energy loss of heavy ions in plasma},
journal = {Nucl. Instrum. Methods Phys. Res. B},
volume = {197},
number = {1},
pages = {22-38},
year = {2002},
issn = {0168-583X},
doi = {https://doi.org/10.1016/S0168-583X(02)01474-X},
url = {https://www.sciencedirect.com/science/article/pii/S0168583X0201474X},
author = {G\"unter Zwicknagel}
}

@article{Montanari2017,
  title = {Low- and intermediate-energy stopping power of protons and antiprotons in solid targets},
  author = {Montanari, C. C. and Miraglia, J. E.},
  journal = {Phys. Rev. A},
  volume = {96},
  issue = {1},
  pages = {012707},
  numpages = {12},
  year = {2017},
  month = {Jul},
  publisher = {American Physical Society},
  doi = {10.1103/PhysRevA.96.012707},
  url = {https://link.aps.org/doi/10.1103/PhysRevA.96.012707}
}

@article{Moldabekov2020,
  title = {Ion energy-loss characteristics and friction in a free-electron gas at warm dense matter and nonideal dense plasma conditions},
  author = {Moldabekov, Zh. A. and Dornheim, T. and Bonitz, M. and Ramazanov, T. S.},
  journal = {Phys. Rev. E},
  volume = {101},
  issue = {5},
  pages = {053203},
  numpages = {14},
  year = {2020},
  month = {May},
  publisher = {American Physical Society},
  doi = {10.1103/PhysRevE.101.053203},
  url = {https://link.aps.org/doi/10.1103/PhysRevE.101.053203}
}

@article{Dai2010,
  title = {Unified First Principles Description from Warm Dense Matter to Ideal Ionized Gas Plasma: Electron-Ion Collisions Induced Friction},
  author = {Dai, Jiayu and Hou, Yong and Yuan, Jianmin},
  journal = {Phys. Rev. Lett.},
  volume = {104},
  issue = {24},
  pages = {245001},
  numpages = {4},
  year = {2010},
  month = {Jun},
  publisher = {American Physical Society},
  doi = {10.1103/PhysRevLett.104.245001},
  url = {https://link.aps.org/doi/10.1103/PhysRevLett.104.245001}
}

@article{Fu2018,
    author = {Fu, Yongsheng and Hou, Yong and Kang, Dongdong and Gao, Cheng and Jin, Fengtao and Yuan, Jianmin},
    title = {Multi-charge-state molecular dynamics and self-diffusion coefficient in the warm dense matter regime},
    journal = {Phys. Plasmas},
    volume = {25},
    number = {1},
    pages = {012701},
    year = {2018},
    month = {01},
    doi = {10.1063/1.5000757},
    url = {https://doi.org/10.1063/1.5000757}
}

@article{Hou2021,
    author = {Hou, Yong and Jin, Yang and Zhang, Ping and Kang, Dongdong and Gao, Cheng and Redmer, Ronald and Yuan, Jianmin},
    title = {Ionic self-diffusion coefficient and shear viscosity of high-Z materials in the hot dense regime},
    journal = {Matter Radiat. Extrem.},
    volume = {6},
    number = {2},
    pages = {026901},
    year = {2021},
    month = {01},
    doi = {10.1063/5.0024409},
    url = {https://doi.org/10.1063/5.0024409}
}

@article{Hansen1973,
  title = {Statistical Mechanics of Dense Ionized Matter. I. Equilibrium Properties of the Classical One-Component Plasma},
  author = {Hansen, Jean Pierre},
  journal = {Phys. Rev. A},
  volume = {8},
  issue = {6},
  pages = {3096--3109},
  numpages = {0},
  year = {1973},
  month = {Dec},
  publisher = {American Physical Society},
  doi = {10.1103/PhysRevA.8.3096},
  url = {https://link.aps.org/doi/10.1103/PhysRevA.8.3096}
}

@article{Tanaka2017,
author = {Tanaka, Shigenori},
title = {Improved equation of state for finite-temperature spin-polarized electron liquids on the basis of Singwi–Tosi–Land–Sjölander approximation},
journal = {Contrib. Plasma Phys.},
volume = {57},
number = {3},
pages = {126-136},
doi = {https://doi.org/10.1002/ctpp.201600096},
url = {https://onlinelibrary.wiley.com/doi/abs/10.1002/ctpp.201600096},
year = {2017}
}

@article{Dandrea1986,
  title = {Electron liquid at any degeneracy},
  author = {Dandrea, R. G. and Ashcroft, N. W. and Carlsson, A. E.},
  journal = {Phys. Rev. B},
  volume = {34},
  issue = {4},
  pages = {2097--2111},
  numpages = {0},
  year = {1986},
  month = {Aug},
  publisher = {American Physical Society},
  doi = {10.1103/PhysRevB.34.2097},
  url = {https://link.aps.org/doi/10.1103/PhysRevB.34.2097}
}

@article{Arista1984,
  title = {Dielectric response of quantum plasmas in thermal equilibrium},
  author = {Arista, N\'estor R. and Brandt, Werner},
  journal = {Phys. Rev. A},
  volume = {29},
  issue = {3},
  pages = {1471--1480},
  numpages = {0},
  year = {1984},
  month = {Mar},
  publisher = {American Physical Society},
  doi = {10.1103/PhysRevA.29.1471},
  url = {https://link.aps.org/doi/10.1103/PhysRevA.29.1471}
}

@article{Arista1981,
  title = {Energy loss and straggling of charged particles in plasmas of all degeneracies},
  author = {Arista, N\'estor R. and Brandt, Werner},
  journal = {Phys. Rev. A},
  volume = {23},
  issue = {4},
  pages = {1898--1905},
  numpages = {0},
  year = {1981},
  month = {Apr},
  publisher = {American Physical Society},
  doi = {10.1103/PhysRevA.23.1898},
  url = {https://link.aps.org/doi/10.1103/PhysRevA.23.1898}
}

@article{Ara2021,
    author = {Ara, J. and Coloma, Ll. and Tkachenko, I. M.},
    title = {Static properties of a warm dense uniform electron gas},
    journal = {Phys. Plasmas},
    volume = {28},
    number = {11},
    pages = {112704},
    year = {2021},
    month = {11},
    doi = {10.1063/5.0062259},
    url = {https://doi.org/10.1063/5.0062259}
}

@article{Kugler1970,
  title = {Bounds for Some Equilibrium Properties of an Electron Gas},
  author = {Kugler, Alfred A.},
  journal = {Phys. Rev. A},
  volume = {1},
  issue = {6},
  pages = {1688--1696},
  numpages = {0},
  year = {1970},
  month = {Jun},
  publisher = {American Physical Society},
  doi = {10.1103/PhysRevA.1.1688},
  url = {https://link.aps.org/doi/10.1103/PhysRevA.1.1688}
}

@article{Kimball1973,
  title = {Short-Range Correlations and Electron-Gas Response Functions},
  author = {Kimball, J. C.},
  journal = {Phys. Rev. A},
  volume = {7},
  issue = {5},
  pages = {1648--1652},
  numpages = {0},
  year = {1973},
  month = {May},
  publisher = {American Physical Society},
  doi = {10.1103/PhysRevA.7.1648},
  url = {https://link.aps.org/doi/10.1103/PhysRevA.7.1648}
}

@article{Bretonnet1988,
  title = {Analytic form for the one-component plasma structure factor},
  author = {Bretonnet, Jean-Louis and Derouiche, Abdelali},
  journal = {Phys. Rev. B},
  volume = {38},
  issue = {13},
  pages = {9255--9256},
  numpages = {0},
  year = {1988},
  month = {Nov},
  publisher = {American Physical Society},
  doi = {10.1103/PhysRevB.38.9255},
  url = {https://link.aps.org/doi/10.1103/PhysRevB.38.9255}
}

@article{Veysman2016,
  title = {Optical conductivity of warm dense matter within a wide frequency range using quantum statistical and kinetic approaches},
  author = {Veysman, M. and R\"opke, G. and Winkel, M. and Reinholz, H.},
  journal = {Phys. Rev. E},
  volume = {94},
  issue = {1},
  pages = {013203},
  numpages = {23},
  year = {2016},
  month = {Jul},
  publisher = {American Physical Society},
  doi = {10.1103/PhysRevE.94.013203},
  url = {https://link.aps.org/doi/10.1103/PhysRevE.94.013203}
}

@article{Tanaka1985,
author = {Tanaka ,Shigenori and Ichimaru ,Setsuo},
title = {Stopping Power of Degenerate Electron Liquid at Metallic Densities},
journal = {J. Phys. Soc. Jpn.},
volume = {54},
number = {7},
pages = {2537-2542},
year = {1985},
doi = {10.1143/JPSJ.54.2537},
URL = {https://doi.org/10.1143/JPSJ.54.2537}
}

@article{Bringa1996,
  title = {Energy loss of correlated ions in plasmas: Collective and individual contributions},
  author = {Bringa, E. M. and Arista, N. R.},
  journal = {Phys. Rev. E},
  volume = {54},
  issue = {4},
  pages = {4101--4111},
  numpages = {0},
  year = {1996},
  month = {Oct},
  publisher = {American Physical Society},
  doi = {10.1103/PhysRevE.54.4101},
  url = {https://link.aps.org/doi/10.1103/PhysRevE.54.4101}
}

@article{Archubi2022,
  title = {General formulation of interactions and energy loss of particles in plasmas: Quantum-wave-packet model versus a semiclassical approach},
  author = {Archubi, C. D. and Arista, N. R.},
  journal = {Phys. Rev. A},
  volume = {105},
  issue = {3},
  pages = {032806},
  numpages = {20},
  year = {2022},
  month = {Mar},
  publisher = {American Physical Society},
  doi = {10.1103/PhysRevA.105.032806},
  url = {https://link.aps.org/doi/10.1103/PhysRevA.105.032806}
}

@article{Arkhipov2019,
author = {Arkhipov, Yu.V. and Ashikbayeva, A.B. and Askaruly, A. and Dubovtsev, D.Yu. and Syzganbayeva, S.A. and Tkachenko, I.M.},
title = {Stopping power of an electron gas: The sum rule approach},
journal = {Contrib. Plasma Phys.},
volume = {59},
number = {6},
pages = {e201800171},
doi = {https://doi.org/10.1002/ctpp.201800171},
url = {https://onlinelibrary.wiley.com/doi/abs/10.1002/ctpp.201800171},
year = {2019}
}

@article{Tolias2023,
    author = {Tolias, Panagiotis and Lucco Castello, Federico and Dornheim, Tobias},
    title = {Quantum version of the integral equation theory-based dielectric scheme for strongly coupled electron liquids},
    journal = {J. Chem. Phys.},
    volume = {158},
    number = {14},
    pages = {141102},
    year = {2023},
    month = {04},
    doi = {10.1063/5.0145687},
    url = {https://doi.org/10.1063/5.0145687}
}

@article{Singh1983PRA,
  title = {Structure factor of liquid alkali metals},
  author = {Singh, H. B. and Holz, A.},
  journal = {Phys. Rev. A},
  volume = {28},
  issue = {2},
  pages = {1108--1113},
  numpages = {0},
  year = {1983},
  month = {Aug},
  publisher = {American Physical Society},
  doi = {10.1103/PhysRevA.28.1108},
  url = {https://link.aps.org/doi/10.1103/PhysRevA.28.1108}
}

@article{Singh1983,
  title = {Structure and thermodynamics of the classical one-component plasma},
  author = {Singh, H. B.},
  journal = {J. Stat. Phys.},
  volume = {33},
  pages = {371–383},
  year = {1983},
  doi = {10.1007/BF01009802},
  url = {https://doi.org/10.1007/BF01009802}
}

@article{Dornheim2022,
title = {The uniform electron gas at high temperatures: ab initio path integral Monte Carlo simulations and analytical theory},
journal = {High Energy Density Phys.},
volume = {45},
pages = {101015},
year = {2022},
issn = {1574-1818},
doi = {https://doi.org/10.1016/j.hedp.2022.101015},
url = {https://www.sciencedirect.com/science/article/pii/S1574181822000386},
author = {Tobias Dornheim and Jan Vorberger and Zhandos Moldabekov and Gerd R\"opke and Wolf-Dietrich Kraeft}
}

@article{Arista1985JPC,
doi = {10.1088/0022-3719/18/26/023},
url = {https://dx.doi.org/10.1088/0022-3719/18/26/023},
year = {1985},
month = {sep},
publisher = {},
volume = {18},
number = {26},
pages = {5127},
author = {N R Arista},
title = {Low-velocity stopping power of semidegenerate quantum plasmas},
journal = {J. Phys. C: Solid State Phys.}
}

@article{Mabey2017,
    author = {Mabey, P. and Richardson, S. and White, T. and Fletcher, L. B. and Glenzer, S. H. and Hartley, N. J. and Vorberger, J. and Gericke, D. O.  and Gregori, G.},
    title = {A strong diffusive ion mode in dense ionized matter predicted by Langevin dynamics},
    journal = {Nat. Commun.},
    volume = {8},
    pages = {14125},
    year = {2017},
    doi = {https://doi.org/10.1038/ncomms14125}
}

@article{Kaehlert2019,
    author = {K\"ahlert, Hanno},
    title = {Dynamic structure factor of strongly coupled Yukawa plasmas with dissipation},
    journal = {Phys. Plasmas},
    volume = {26},
    number = {6},
    pages = {063703},
    year = {2019},
    month = {06},
    doi = {10.1063/1.5099579},
    url = {https://doi.org/10.1063/1.5099579},
}

\end{document}